\documentclass[11pt]{article}
\usepackage{amssymb}
\usepackage{amsmath}
\usepackage{graphics}
\usepackage{color}

\newcounter{linenumber}

\usepackage{times}

\newtheorem{theorem}{Theorem}[section]
\newtheorem{lemma}[theorem]{Lemma}

\newtheorem{proposition}[theorem]{Proposition}
\newcommand{\sq}{\hbox{\rlap{$\sqcap$}$\sqcup$}}
\newcommand{\qed}{\hspace*{\fill}\sq}
\newenvironment{proof}{\noindent {\bf Proof.}\ }{\qed\par\vskip 4mm\par}
\newenvironment{sketch}{\noindent {\bf Sketch of Proof.}\ }{\qed\par\vskip 4mm\par}

\usepackage[ruled]{algorithm}
\usepackage{algpseudocode}
\usepackage{dsfont}
\usepackage{wide}
\alglanguage{pseudocode}

\alglanguage{pseudocode}
\algblockdefx{Procedure}{EndProcedure}[1]{{\bf procedure} #1}{{\bf end procedure}}
\algblockdefx{RepeatForever}{EndRepeatForever}{{\bf repeat forever}}{{\bf end repeat}}

\newcommand{\Rem}[1]{\qquad $\slash *$ #1 $* \slash$}
\newcommand{\X}{\mathcal{X}}
\newcommand{\send}[2]{{\tt send}$\langle$#1$\rangle$\ {\bf to} #2}
\newcommand{\receive}[2]{{\tt receive}$\langle$#1$\rangle$\ {\bf from} #2}
\newcommand{\brd}[1]{{\tt rbroadcast}$\langle$#1$\rangle$}
\newcommand{\deliv}[1]{{\tt rdeliver}$\langle$#1$\rangle$}
\newcommand{\settimer}[1]{{\tt settimer}(#1)}
\newcommand{\expire}[1]{{\tt timerexpire}(#1)}
\newcommand{\N}{\mathds{N}}
\begin{document}


\title{Algorithms For Extracting Timeliness Graphs\thanks{This work has been supported in part by the ANR projet \emph{SHAMAN}.}
}

\author{Carole Delporte-Gallet \\
{\small University Paris Diderot} \\
{\small Carole.Delporte@liafa.jussieu.fr} \\
    \and
St\'ephane Devismes  \\
{\small University Joseph Fourier (Grenoble)}\\
{\small Stephane.Devismes@imag.fr}
   \and
   Hugues Fauconnier\\
{\small University Paris Diderot} \\
{\small Hugues.Fauconnier@liafa.jussieu.fr} \\
  \and
Mikel Larrea\\
{\small University of the Basque Country}\\
{\small Mikel.Larrea@ehu.es}
       }

\date{}

\maketitle \thispagestyle{empty}


\begin{abstract}
  We consider asynchronous message-passing systems in which some links
  are timely and processes may crash. Each run defines a
  \emph{timeliness graph} among correct processes: $(p,q)$ is an
  edge of the timeliness graph if the link from $p$ to $q$ is timely
  (that is, there is bound on communication delays from $p$ to
  $q$). The main goal of this paper is to approximate this timeliness
  graph by graphs having some properties (such as being trees,
  rings,... ).  Given a family $S$ of graphs, for runs such that the
  timeliness graph contains at least one graph in $S$ then using an
  \emph{extraction algorithm}, each correct process has to converge to
  the same graph in $S$ that is, in a precise sense, an approximation
  of the timeliness graph of the run. For example, if the timeliness
  graph contains a ring, then using an extraction algorithm, all
  correct processes eventually converge to the same ring and in this ring
  all nodes will be correct processes and all links will be timely.
  
  We first present a general extraction algorithm and then a more
  specific extraction algorithm that is communication efficient
  (\emph{i.e.}, eventually all the messages of the extraction algorithm use only
  links of the extracted graph).
\end{abstract}

\section{Introduction}

Designing fault-tolerant protocols for asynchronous systems is highly desirable but also highly complex. Some classical agreement problems 
such as {\em consensus} and {\em reliable broadcast} are well-known tools for solving more sophisticated tasks 
in  faulty environments (e.g., \cite{GS01,GL03}). 
Roughly speaking, with  consensus processes must reach a common decision on their inputs, and with reliable broadcast processes must deliver the same set of messages.

It is well known that consensus cannot be solved in asynchronous systems with failures~\cite{FLP85}, and several mechanisms were introduced to circumvent this impossibility result:
 {\em randomization} \cite{CC85}, {\em partial synchrony}~\cite{DDS87,DLS88} and {\em (unreliable) failure detectors}~\cite{CT96}. 
 
Informally, a failure detector is a distributed oracle that gives (possibly incorrect) hints about the process crashes. 
Each process can access a local failure detector module that monitors the processes of the system and maintains a list of 
processes that are suspected of having crashed.

Several classes of failure detectors have been introduced, {\em e.g.}, $\mathcal P$, $\mathcal S$, $\Omega$, etc. 
Failure detectors classes can be compared by reduction algorithms, so for any given problem $P$, a natural question is 
``\emph{What is the weakest failure detector (class) that can solve $P$ ?}''. 
This question has been extensively studied for several problems in systems {\em with infinite process memory}
({\em e.g.}, uniform and non-uniform versions of consensus~\cite{CHT96,EHT07},
non-blocking atomic commit~\cite{DFGHKT04}, uniform reliable broadcast~\cite{AKD99,HR99},
	implementing an atomic register
	in a message-passing system~\cite{DFGHKT04},
	mutual exclusion~\cite{DFGK05},
	boosting obstruction-freedom~\cite{GKK06},
	set consensus~\cite{RT06,PK07}, etc.).
This question, however, has not been as extensively studied in the context of systems {\em with finite process memory}.

In this paper, we consider systems where processes have finite memory,  processes can crash and links can lose messages 
(more precisely, links are fair lossy and 
FIFO\footnote{
  The FIFO assumption is necessary because, from the results in~\cite{lynch88data}, if lossy links are not FIFO, 
  reliable broadcast requires unbounded message headers.
}).
Such environments can be found in many systems, for example in
sensor networks, sensors are typically equipped with small memories, they can crash 
when their batteries run out, and they can experience message losses if they use wireless communication.

In such systems, we consider (the uniform versions of) reliable broadcast, consensus and repeated consensus. Our contribution is threefold: 
First, we establish that the weakest failure detector for reliable broadcast is ${\mathcal P}^-$ --- 
a failure detector that is almost as powerful than the perfect failure detector $\cal P$.\footnote{Note that $\mathcal P \subseteq \mathcal P^-$ and $\mathcal P^-$ is \emph{unrealistic} according to the definition in \cite{DFG02}.}
Next, we show that consensus can be solved using failure detector $\mathcal S$.
Finally, we prove that ${\mathcal P}^-$ is the weakest failure detector for repeated consensus.
Since $\mathcal S$  is strictly weaker than ${\mathcal P}^-$, in some precise sense these results imply that,
in the systems that we consider here, consensus is easier to solve than reliable broadcast, and reliable broadcast  is as difficult to solve as repeated consensus.

The above results are somewhat surprising because, when processes have infinite memory,
reliable broadcast is easier to solve than consensus\footnote{With infinite memory and fair lossy links, (uniform) reliable broadcast can be solved using $ \Theta$~\cite{BN92}, and $\Theta$ is strictly weaker than $(\Sigma, \Omega)$ which is
necessary to solve consensus.}, and  repeated consensus is not more difficult to solve than consensus.

\paragraph{Roadmap.} 
The rest of the paper is organized as follows: In the next section, we present the model considered in this paper. 
In Section \ref{sect:rb}, we show that in case of process memory limitation and possibility of crashes, 
$\mathcal{P}^-$ is necessary and sufficient to solve reliable broadcast. In Section \ref{sect:c}, we show that 
consensus can be solved using a failire detector of type $\mathcal S$ in our systems. In Section \ref{sect:rc}, 
we show that $\mathcal{P}^-$ is necessary and sufficient to solve repeated consensus in this context.

For space considerations, all the proofs are relegated to an optional appendix.

\section{Informal Model}\label{sect:model}

\paragraph{Graphs.}

We begin with some definitions and notations concerning
graphs.  For a directed graph $G=\langle N,E\rangle$, $Node(G)$ and
$Edge(G)$ denote $N$ and $E$, respectively.  Given a graph $G$ and a
set $M\subseteq Node(G)$, $G[M]$ is the {\em subgraph} of $G$ induced
by $M$, {\em i.e.}, $G[M]$ is the graph $\langle M,Edge(G)[M]\rangle$
where $(p,q) \in Edge(G)[M] $ if and only if $p,q \in M$ and $(p,q)
\in Edge(G)$.

The tuple $(X,Y)$ is a \emph{directed cut} (\emph{dicut} for short) of
$G$ if and only if $X$ and $Y$ define a partition of
$Node(G)$
 and there is no
directed edge $(y,x)\in Edge(G)$ such that $x\in X$ and $y\in Y$. We
say that $G'$ is a \emph{dicut reduction} from $G$ if there exists a
dicut $(X,Y)$ of $G$ such that $G'=G[X]$.  A set $S$ of graphs is
\emph{dicut-closed} if and only if it is closed under dicut reduction,
namely if $G\in S$ then all the graphs obtained by a dicut-reduction
of $G$ are in $S$.



\paragraph{Processes and Links.}

We consider distributed systems composed of $n$ processes which communicate by
message-passing through directed links. We denote the set of processes
by $\Pi=\{p_1,...,p_n\}$. We assume that the communication graph is complete,
{\em i.e.}, for each pair of distinct processes $(p,q)$,
there is a directed link from $p$ to $q$.

A process may fail by crashing, in which case it definitively stops
its local algorithm. A process that never crashes is said to be
\emph{correct}, \emph{faulty} otherwise.

The (directed) links are {\em reliable}, {\em i.e.} every message sent
through a link $(p,q)$ is eventually received by $q$ if $q$ is correct
and  if a message $m$ from $p$
is received by $q$, $m$ is received by $q$ at most once, and only if
$p$ previously sent $m$ to $q$.

The links being reliable, an implementation of the {\em reliable
  broadcast}~\cite{HT94} is possible. A reliable broadcast is defined
with two primitives: \brd{$\emph m$} and \deliv{$\emph
  m$}. Informally, after a correct process $p$ invokes \brd{$\emph m$}, all
correct processes eventually  \deliv{$\emph m$};
 after a faulty process $p$ invokes \brd{$\emph m$}, either  all
correct processes eventually  \deliv{$\emph m$} or 
correct processes never  \deliv{$\emph m$}.

\paragraph{Timeliness.}

To simplify the presentation, we assume the existence of a discrete
global clock. This is merely a fictional device:  
the processes do not have access to it. We take the range $\mathcal T$
of the clock's ticks to be the set of natural numbers. 

We assume that every correct process $p$ is {\em timely}, {\em i.e.},
there is a lower and an upper bound on the execution rate of
$p$. Correct processes also have clocks that are not necessarily
synchronized but we assume that they can accurately measure intervals
of time.

A link $(p,q)$ is \emph{timely} if there is an unknown bound
$\delta$  such that no message sent by $p$ to $q$ at
time $t$ may be received by $q$ after time $t+\delta$.

A \emph{timeliness graph} is simply a directed graph whose set of 
nodes are a
subset of $\Pi$. The timeliness graph represents the timeliness
properties of the links. 
Intuitively, for timeliness graph $G$, $Node(G)$  is the set of correct
processes and $(p,q)$ is in $Edge(G)$ if and only
if the link $(p,q)$ is timely.

\paragraph{Runs.}
An algorithm $\cal A$ consists of $n$ deterministic (infinite)
automata, one for each process; the automaton for process $p$ is
denoted ${\cal A}(p)$. The execution of an algorithm $\cal A$
proceeds as a sequence of process \emph{steps}. Each process performs its steps atomically. During a step, a process may send and/or receive some messages and changes its state.

A run $r$ of algorithm $\cal A$ is a tuple $r =\langle T, I , E,
S\rangle $ where $T$ is 
a timeliness graph, $I$ is the initial state of the processes
in $\Pi$, $E$ is an infinite sequence of steps of $\cal A$, and $S$ is
a list of increasing time values indicating when each step in $E$
occurred. 
A run must satisfy usual properties concerning sending and receiving
messages.
Moreover, we assume that (1) all correct processes make an infinite number
of steps: $p\in Node(G)$ if and only if $p$ makes an infinite number of
steps in $E$ and (2) the timeliness of links is deduced from the timeliness graph:
$(p,q)\in Edge(G)$ if and only if the link $(p,q)$
is timely in $E$.

In the following for run $r =\langle T, I , E,
S\rangle $, $T(r)$ denotes $T$ the timeliness graph of $r$, and
$Correct(r)$ is the set of correct processes for the run $r$, namely,
$Correct(r)=Node(T(r))$. Note that by definition, $(p,q)$ is a timely
link if and only if $(p,q)\in Edge(T)$.

Remark that in the definition given here a link may be timely even if
no message is sent on the link.  If link $(p,q)$ is FIFO ({\em i.e.},
messages from $p$ to $q$ are received in the order they are sent) and
$p$ regularly sends messages to $q$, then the timeliness of these
messages implies the timeliness of the link itself. So in the
following we always assume that links are FIFO.

\subsection{Some Systems}

We say that timeliness graph $G$ is \emph{compatible with timeliness
  graph} $G'$ if and only if (1) $Node(G)=Node(G')$ and (2) $Edge(G)
\subseteq Edge(G')$.  By extension, timeliness graph $G$ is
\emph{compatible with run} $r$ if $G$ is compatible with $T(r)$, the
timeliness graph of $r$.
Hence, timeliness graph  $G$ is compatible with run $r$
if $Node(G)$ is the set of correct processes in $r$ and if $(p,q)$ is
an edge of $G$ then
 $(p,q)$ is timely in $r$.

\vspace{0.5em}

A \emph{system} $\cal X$ is defined as a set of  timeliness
graphs. 
The set of runs of system $\cal X$ denoted $R({\cal X})$ is the set of
all runs $r$ such that there exists a timeliness graph $G$ in
$\cal X$ compatible with $r$.

Below, we define the systems considered in this paper:
\begin{itemize}
\item
$\cal ASYNC$ is the set of all timeliness graphs $G$ such that
$Edge(G)=\emptyset$. In $\cal ASYNC$ there is no timeliness
assumption about links and $R({\cal ASYNC})$ is the set of
all runs in an asynchronous system.
\item
$\cal COMPLETE$ is the set of all complete graphs whose nodes
are the subsets of $\Pi$.

\item $\cal STAR$ is the set of all timeliness graphs with a {\em
    source}, {\em i.e.}, $G \in \cal STAR$ if and only if $Node(G)
  \subseteq \Pi$ and there exists $p_0 \in Node(G)$ (the center of the star
  or the source) such that
 $ Edge(G)=\{(p_0,q)| q \in Node(G)\setminus \{p_0\} \}$. 
 Clearly a run $r$ is in $R({\cal STAR})$ if and only if
  there is at least one {\em source} in $r$.

\item $\cal TREE$ is the set of all timeliness graphs $G$ that
  are rooted directed trees, {\em i.e.}, $|Edge(G)| = |Node(G)|-1$ and
  there exists $p_0$ in $Node(G)$ such that $\forall q \in Node(G)$,
  there is a directed path of $G$ from $p_0$ to $q$. Clearly a run $r$
  is in $R({\cal TREE})$ if and only if there is at least one timely
  path from a correct process to all correct processes.

\item $\cal RING$ is the set of all timeliness graphs $G$ such
 that $G$ is a directed cycle (a ring). Clearly  a run $r$ is in $R({\cal
  RING})$ if and only if there is a timely (directed) cycle over all
  correct processes.

\item $\cal SC$ is the set of all timeliness graphs that are
  strongly connected. Clearly, a run $r$ is in $R({\cal SC})$ if and
  only if there exists a (directed) timely path between each pair of
  distinct correct processes.

\item $\cal BIC$ is the set of all timeliness graphs $G$ such
  that for all $p$, $q \in Node(G)$, there exist at least two distinct
  paths from $p$ to $q$. $\cal BIC$ corresponds to the set of
  2-strongly-connected graphs. Clearly, a run $r$ is in $R({\cal
    BIC})$ if and only if there exists at least two distinct timely
  paths between each pair of distinct correct processes.

\item $\cal PAIR$ is the set of all timeliness graphs $G$ such that
  $Edge(G)=\{(p_0,p_1),$ $(p_1,p_0)\}$ with $p_0,p_1\in Node(G)$ and
  $p_1\neq p_0$. Clearly, a run $r$ is in $R({\cal PAIR})$ if and only
  if there exists two distinct correct processes $p_0$ and $p_1$ such
  that $(p_0,p_1)$ and $(p_1,p_0)$ are timely links.
\end{itemize}

\section{Extraction Algorithms}\label{sect:extract}

Given a system $\cal X$, the goal of an \emph{extraction algorithm}
is to ensure that in each run $r$
in $\cal X$, all correct processes eventually agree on the same
element of $\cal X$ and that this element is, in some precise sense, an
approximation of the timeliness graph of run $r$.

For example, in $\cal RING$,
all processes have to eventually agree on some ring and this
ring has to be compatible with the timeliness graph of the
run. In particular this ring contains all the correct processes.
However, the compatibility relation may be too strong: 
In many systems, it is not possible to distinguish
between a crashed process and a correct one, so the graph $G$ on which
the processes eventually agree may contain crashed
processes and then the graph is not exactly compatible with the run.
Then we weaken the compatibility and impose only that
the subgraph of $G$ induced by
the set of correct processes of the run is a dicut reduction of the
timeliness graph of the run.

We now formally define what an extraction algorithm is. First, in such
an algorithm, every process $p$ maintains a local variable $G_p$ which
contains a timeliness graph. Then, we say that an
algorithm {\em extracts a timeliness graph in ${\cal X}$} if and only
if for every run $r$ in ${\cal X}$ there is a timeliness graph $G$ (called
the \emph{extracted graph}) such
that:
\begin{itemize}
\item \emph{Convergence:} for all correct processes $p$ there is a
  time $t$ after which $G_p=G$
\item \emph{Compatibility:} $G[Correct(r)]$ is compatible with $T(r)$
\item \emph{Closure:} $G[Correct(r)]$ is a dicut reduction of $G$ or
  is equal to $G$
\item \emph{Validity:} $G$ is in $\cal X$
\end{itemize}

Remark that for all systems that contain ${\cal ASYNC}$ there is a trivial
extraction algorithm: for each run processes extract the graph $G$
such that $Node(G)=\Pi$ and $Edge(G)=\emptyset$.





A more constrained version of the extraction problem is the following:
an algorithm $\cal A$ \emph{extracts exactly} timeliness graphs in $\cal X$ if
for every run $r$ in system $\cal X$, the extracted graph $G$ is
compatible with $T(r)$. In this case, all correct processes eventually
know the exact set of correct processes: it is the set of nodes of
the extracted graph.


\paragraph{Some Results about Extraction Algorithms.}


First we show that an extraction algorithm may help to route messages
using only timely links:
\begin{lemma}\label{cut}
  Let $G$ be a graph extracted from run $r$, if $(p,q)$ is in
  $Edge(G)$ and $q$ is a correct process then $p$ is correct.
\end{lemma}
\begin{proof}
  By contradiction, assume that $p$ is not correct, then 
  $(Correct(r), Node(G)-Correct(r))$ is not a dicut because $(p,q)\in
  Edge(G)$, $p\in Node(G)-Correct(r)$ and $q\in Correct(r)$, which
  contradicts the Closure property.
\end{proof}

From this lemma and the Compatibility property, we deduce directly:
\begin{proposition}\label{route}
If $(p=p_0, \ldots,p_i, \ldots , q=p_m)$ is a path in the extracted
graph and $p$ and 
$q$ are  correct processes, then for every $i$ such that  $0\leq i <
m$  the link
$(p_i,p_{i+1})$ is timely and process $p_i$ is correct.
\end{proposition}

From a practical point of view, this proposition shows that the
extracted graph may be used to route messages between processes using
only timely links: the route from $p$ to $q$ is a path in the
extracted graph (if any). All intermediate nodes are correct processes
and agree on the extracted graph and then on the path.

For example with
$\cal TREE$, the tree extracted by the algorithm enables to route
messages from the root of the tree to any other processes and the
routing uses only timely links.

\vspace{0.5em}

Generally, the main goal of the extraction algorithm
is not only to extract a graph $G$ in ${\cal X}$ but also to ensure
that 
$G[Correct(r)]$ is in ${\cal X}$ (even if the processes do not know the
set of correct processes).  In particular, this property is ensured if
${\cal X}$ is dicut-closed: 
the Closure property implies that $G[Correct(r)]$ is in ${\cal
  X}$.

Among the systems we consider, only system $\cal PAIR$ is not
dicut-closed: $H=\langle \{x\}, \emptyset\rangle$ is a dicut
reduction of $G=\langle \{x,y,z\}, \{(y,z),(z,y)\}\rangle$ but is not in
$\cal PAIR$. It is easy to verify that every other previously
introduced system is dicut-closed. For these systems we obtain:
\begin{proposition}\label{prop:1}
Consider any extraction algorithm for the system ${\cal X}$.
\begin{itemize}
\item If ${\cal X} = {\cal STAR}$, then the center of the extracted
  star is a correct process.
\item If ${\cal X} = {\cal TREE}$, then the root of the extracted tree
  is a correct process.  
\item If ${\cal X} \in \{{  {\cal SC}, \cal COMPLETE}, {\cal RING}, {\cal
    BIC}\}$, then the extraction is exact.
\end{itemize}
\end{proposition}
\begin{proof}
  For $\cal STAR$ and $\cal TREE$, all the dicut reductions of the
  extracted graph contain at least respectively the center and the
  root, then the restriction of the extracted graph contains at least
  these nodes, proving that they are correct processes.
  
  There is no dicut for a strongly connected graph. Hence
in $\cal SC$, there is
  no dicut reduction then by the Closure property the subgraph induced by the
  set of correct processes of the extracted graph is the extracted
  graph itself.
  $\cal COMPLETE$, $\cal RING$, and $\cal BIC$ are
  particular cases of systems only composed of strongly connected
  timeliness graphs.
\end{proof}

An immediate consequence of Proposition \ref{prop:1} is that any
extraction algorithm gives an implementation of eventual leader
election (failure detector $\Omega$) for systems $\cal STAR$ and $\cal
TREE$ as well as an implementation of failure detector $\Diamond \cal
P$ for systems $\cal COMPLETE$, $\cal RING$, $\cal SC$ and $\cal BIC$.


%
\vspace{0.5em}

Due to the lack of space, the proofs of the two following propositions
have been moved in the appendix. In the first proposition we show that
extraction is not always possible. Actually, in the proof we exhibit
some non dicut-closed systems, namely ${\cal PAIR}$, where no
extraction algorithm can be implemented.


\begin{proposition}\label{prop:noalgo}
  There exist some systems $\cal X$ for which there
  is no extraction algorithm.
\end{proposition}

 In the next section we show that for all
dicut-closed systems there is an extraction algorithm. 
For systems like ${\cal STAR}$,  ${\cal TREE}$ and  ${\cal PAIR}$, there exists no {\em exact} extraction algorithm.

\begin{proposition}\label{prop:noexact}
  There exist some systems $\cal X$ for which there
  is an  extraction algorithm and there is no exact extraction algorithm.
\end{proposition}


\section{An Extraction Algorithm}\label{sect:brut}

The aim of this section is to show that the dicut-closed property of a
system is sufficient to solve the extraction problem. To that end, we
propose in Figure~\ref{algo:brut} an extraction algorithm, called
${\cal A}({\cal X})$, for dicut-closed systems $\cal X$. 

The basic idea of Algorithm ${\cal A}({\cal X})$ is to make processes
select a graph that is compatible with  the timeliness graph
of the run.  For this, each process maintains for each graph $x$
in $\cal X$ an {\em accusation counter} $Acc[x]$. This counter
infinitely grows if some correct process is not in $x$ or if some
directed edge of $x$ is not timely.  Then, $Acc[x]$ is bounded if and
only if $x$ contains all correct processes and all  timely
links between pairs of correct processes.

We implement accusation counters as follows. 
A process regularly blames all the graphs in ${\cal X}$ in which it is
not a node: it
increments the accusation counters of all these 
graphs. 
Note that if the process is correct this accusation is
justified  and if the process is not correct, after some time, the process
being dead stops to increment the accusation counters.  
Moreover,
each process
regularly sends on its outgoing links $alive$ messages. 
Each process 
maintains an estimate of the communication delays for each incoming
link ($\Delta[q]$ for the incoming link $(q,p)$). 
If it does not
receive $alive$ messages within these estimates on some incoming link it blames
all timeliness graphs in ${\cal X}$ containing this link (\emph{i.e.},
increments the accusation counters for these graphs).
As the estimate of the communication delay may be too short, each time it is
exceeded the process increases it for the link. 
In this way, if the
link is timely, at some time the estimate will be greater than the bound on communication delay. 


The accusation counters are broadcast by reliable broadcasts. Each
time a process receives a new value of accusation counter it updates
its own accusation counter to the maximum of the received values and its
current values.
Hence, if some timely graph stops to be blamed then 
all correct processes eventually agree on the value of its accusation
counter.

By selecting the graph $G$ with the lowest accusation value (to break
ties, we assume a total order among the graphs of $\cal X$) if any, 
correct
processes eventually agree on the same timeliness graph of $\cal X$, moreover we
can prove that this graph
contains (1) all the correct processes, and (2) all edges between
correct processes are timely links.
 As a consequence, the
Convergence, the Compatibility and the Validity properties of the
extraction algorithm are ensured.  Nevertheless, this graph can also contain
faulty processes and edges between correct and faulty processes.



Consider now the Closure property. If $G$ contains only correct
processes then the Closure property is trivially satisfied. Otherwise,
$G$ contains $Correct(r)$ and a set $F$ of faulty processes. In this
case, $(Correct(r),F)$ is a dicut reduction of $G$: Indeed if there is
an edge in $G$ from a faulty process $q$ to a correct process $p$,
eventually the process $p$ stops to receive messages from $q$ and the
accusation counter of $G$ grows infinitively often.  Hence, in all
cases, the Closure property is satisfied.

Hence, if $\cal X$ is dicut-closed, Algorithm
${\cal A}({\X})$ extracts a graph in $\cal X$. Moreover
from Proposition~\ref{prop:1}, if all the graphs  of $\cal X$ are
strongly connected then the algorithm exactly extracts   a
graph in ${\cal X}$.

\vspace{0.5em} 

In the algorithm, each process $p$ uses local timers,
one per process.  The timer of $p$ dedicated to $q$ is set (by
setting \settimer{$q$} to a positive value) to a time interval rather
than absolute time.  The timer is decremented until it expires.  When
the timer expires \expire{$q$} becomes $true$.  Note that a timer can
be restarted before it expires.

In the algorithm, we denote by $ \prec $ the total order relation
on $\cal X$ and by $\prec_{lex}$ (see Line \ref{winner}) the total order relation
defined as follows: $\forall x,y
\in \X$, $\forall c_x, c_y \in \N$, $(c_x,x) \prec_{lex} (c_y,y)
\equiv [c_x < c_y \vee (c_x = c_y \wedge x \prec y)]$.

\vspace{0.5em}

\begin{figure}[htpb]
\scriptsize
{\tt Code for each process $p$}
\begin{algorithmic}[1]
\State {\bf Procedure} $updateExtractedGraph()$
\State \qquad   $G \gets x$ such that $(Acc[x],x) = \min_{\prec_{lex}}\{(Acc[x'],x')$ such that $x' \in \X\}$ \label{winner}

\vspace{1em}

\State {On initialization:}
\State {\bf for all} {$x \in \X$} {\bf do} $Acc[x] \gets 0$


\State {\bf for all} {$q \in \Pi \setminus \{p\}$} {\bf do}
\State \qquad $\Delta[q] \gets 1$
\State \qquad \settimer{$q$} $\gets \Delta[q]$
\State  $updateExtractedGraph()$
\State  {\bf start tasks} 1 and  2



\vspace{1em}

\State {task 1:}
\State  \qquad {\bf loop forever}
\State  \qquad   \qquad \send{$alive$}{every $q \in \Pi \setminus \{p\} $ {\bf every $K$ time}}
\State  \qquad   \qquad \brd{$ACC$,$\perp$,$p$} {\bf every $K$ time}  \Rem{to accuse graphs that do not contain $p$}


\vspace{0.5em}
\State {task 2:}
\State \qquad   {\bf upon} {\receive{$alive$}{$q$}} {\bf do}
\State \qquad  \qquad  \settimer{$q$} $\gets \Delta[q]$ 
\vspace{0.5em}
\State  \qquad {\bf upon} {\expire{$q$}} {\bf do}
\State \qquad  \qquad  \brd{$ACC,q,p$} \Rem{to accuse graphs that contain the link $(q,p)$}
\State \qquad  \qquad \qquad $\Delta[q] \gets \Delta[q]+1$
\State \qquad  \qquad \qquad \settimer{$q$} $\gets \Delta[q]$

\vspace{0.5em}

\State \qquad  {\bf upon} {\deliv{$ACC$,$q$,$h$}} {\bf do}\Rem{information from $h$} 
\State \qquad  \qquad {\bf for all} {$x \in \X$} {\bf do}
\State \qquad  \qquad \qquad {\bf if} {$q = \perp$} {\bf then}
\State \qquad  \qquad \qquad \qquad {\bf if} {$h \notin Node(x)$} {\bf then} $Acc[x] \gets Acc[x] + 1$
\State \qquad  \qquad \qquad {\bf else} 
\State \qquad  \qquad \qquad \qquad {\bf if} {$(q,h) \in Edge(x)$} {\bf then} $Acc[x] \gets Acc[x] + 1$
\State \qquad  \qquad $updateExtractedGraph()$



\end{algorithmic}

\caption{Algorithm ${\cal A}({\X})$ extracts a graph in $\X$ }\label{algo:brut}
\end{figure}

A sketch of the correctness proof of ${\cal A}({\X})$ is given
below. In this sketch, we consider a run $r$ of ${\cal A}({\X})$ in
dicut-closed system $\cal X$.
We will denote by $var_p^t$ the value of $var$ of process $p$ at time $t$.

We first notice that all variables $Acc_p[x]$ are monotonically
increasing:
\begin{lemma}
  For all times $t$ and $t'$ such that $t \ge t'$, for all processes
  $p$, for all graphs $x$ in $\cal X$, $Acc^t_p[x] \ge Acc^{t'}_p[x]
  $.
\end{lemma}
Let $\sup(Acc_p[x])$ be the supremum  of $Acc_p^t[x]$ for
all $t$, we say 
that $Acc_p[x]$ is unbounded if $\sup(Acc_p[x])$ is equal to $\infty$
and bounded otherwise.
As $Acc_p[x]$ is also updated by reliable broadcast each time some
process $q$ modifies $Acc_q[x]$ we have:

\begin{lemma}
  For all  correct processes $p$ and  $q$, for all graphs $x$ in $\cal X$, $\sup(Acc_p[x])$ $=$ $\sup(Acc_q[x])$
\end{lemma}
Let $\sup(Acc[x])$ be  the supremum $\sup(Acc_p[x])$ over all correct process $p$ of $Acc_p[x]$, then
$\sup(Acc[x])$ is well-defined.
If there is a least one $x\in {\cal X}$ such that
$\sup(Acc[x])$ is bounded, then $\min\{\sup(Acc[x]) | x'\in
{\cal X}\}$ is finite, hence $G$ the graph such that
$(Acc[G],G)=\min_{\prec_{lex}}\{ (Acc[x'], x')| x' \in \X\}$ is well defined.
Then all correct processes converge to the same graph:
\begin{lemma}\label{con}
If  there exists $x$ in ${\cal X}$ such that
$\sup(Acc[x])$ is bounded then there is a time after which for every
correct process $p$, $G_p$ is $G$.
\end{lemma}
Now prove the Compatibility property.
Consider any timeliness graph compatible with $T(r)$, 
and assume
that $x\in {\cal X}$, then there is a time $t_0$ after which all faulty
processes are dead and the estimates of communication delays are
greater than the bounds of communication delays of timely links of the
run. After time $t_0$, (1) as $x$ contains all correct processes, no
process will blame $x$ because it is not a node of $x$, and (2) as all
edges of $x$ are timely, no process will blame $x$ for one of its
edges then:

\begin{lemma}\label{exist}
If $x$ in ${\cal X}$ is compatible with $T(r)$, then $\sup(Acc[x])$ is
bounded.
\end{lemma}

Reciprocally, let $x$ be a timeliness graph of $\X$ that is  not compatible with
the run.
If process $p$ is not correct there is a time $t$ after which it does
not send any $alive$ message, and there is a time after the timers on $p$
expire forever for all correct processes, then if $p$ is a node of
some $x\in {\cal 
  X}$, $Acc_p[x]$ is incremented infinitely often and
$\sup(Acc[x])=\infty$. 
In the same way if $(p,q)$ is not
timely, by the fifo property of the link, the timer for $p$ expires infinitely often for process $q$
and if $(p,q)$ is
an edge of $x$  then $Acc_q[x]$ is incremented infinitely
often and $\sup(Acc[x])=\infty$.
 
Then:
\begin{lemma}
For every $x$ in $\X$,  if $\sup(Acc[x])$ is bounded then
$x[Correct(r)]$ is compatible with $T(r)$.
\end{lemma}

Hence:

\begin{lemma}\label{compat}(Compatibility)
$G[Correct(r)]$ is compatible with $T(r)$.
\end{lemma}

It remains to prove that $G$ satisfies the Closure
property: $G[Correct(r)] $ is a dicut reduction of $G$ or is equal to
$G$.
 As $G[Correct(r)]$ is compatible with $T(r)$, we have:
\begin{lemma}\label{d}
$Correct(r)\subseteq Node(G)$.
\end{lemma}
Let $F=Node(G)-Correct(r)$.
If $F$ is empty the Closure property is trivially ensured.  Consider
now the case where $F$ is not empty. $F$ contains only faulty
processes and $(Correct(r),F)$ is a partition of $G(Node)$.
 If there is an edge in $Edge(G)$ from a faulty process $q$
to a correct process $p$, eventually the process $p$ never receives a
message from $q$ and the accusation counter of $G$ will be unbounded,
contradicting the choice of $G$. So, we have:

\begin{lemma}\label{dicutclose}
If  $F \neq \emptyset$ then  $Edge(G) \cap( F\times Correct(r))=\emptyset$.
\end{lemma}
Hence, $(Correct(r),F)$ is a dicut of $G$.

Lemma~\ref{con} and Lemma~\ref{exist} prove the Convergence property,
Lemma~\ref{compat} proves the Compatibility property and
Lemma~\ref{dicutclose} proves the Closure property. Moreover, $G$ is
clearly in ${\cal X}$ proving the Validity.
Proposition~\ref{prop:1} shows that the extraction is exact
when all graphs of $\X$ are strongly connected. Hence, we
can conclude with the following theorem:

\begin{theorem}\label{th:brutCN}
  Let $\X$ be a dicut-closed system. Algorithm ${\cal A}({\X})$
  extracts a graph in $\cal X$.  Moreover if all graphs of $\X$ are
  strongly connected, Algorithm ${\cal A}({\X})$ exactly extracts a
  graph in $\X$.
\end{theorem}


\section{An Efficient Extraction Algorithm}\label{sect:eff}

In this section, we propose another extraction algorithm called ${\cal
  AF}({\X})$ (Figures \ref{algo:eff2} and \ref{algo:eff}). This
algorithm is efficient meaning that the (correct) processes eventually
only send messages along the edges of the extracted graph.

${\cal AF}({\X})$ (exactly) extracts a timeliness graph from system
$\X$, where (1) $\X$ is dicut-closed and (2) for all graphs $g \in \X$
there is some process $p$, called {\em root}, such that there is a
directed path from $p$ to every node of $g$. 
For example, $\cal TREE$  and  $\cal RING$  systems have this property.

In the following, we refer to these systems as {\em dicut-closed
  systems with a root}. For every graph $g$ in $\X$, the function
$root(g)$ returns a root of $g$.


In the algorithm, every process $p$ stores several values concerning
the graphs $x \in \X$ such that $root(x) = p$:
(1) $Acc[x]$ is the accusation counter of $x$ whose goal is the same
  as in Algorithm \ref{algo:brut},
(2) $Prop[x]$ is a {\em proposition counter} whose goal will be
  explained later, and
(3) $\Delta[x]$ gives the expected time for a message to go from $p$
  (the root of the $x$) to  all the nodes of $x$.

Every process also maintains a set variable $Candidates$.  Each
element of this set is a 4-tuple composed of a graph $x$ of $\X$ and
the newest values of $Acc[x]$, $Prop[x]$, and $\Delta[x]$ known by
the process (the exact values are maintained at $root(x)$). Each
element in this set is called {\em candidate} and each process selects
its extracted graph among the graphs in the candidate elements.

As in Algorithm \ref{algo:brut}: 
\begin{itemize}
\item[(1)] Each process $p$ sends $alive$ messages on its outgoing links
  and monitors its incoming links. However, we restrain here the
  $alive$ message sendings: process $p$ sends $alive$ messages on its
  outgoing link $(p,q)$ only if $(p,q)$ is in a graph candidate.
\item[(2)] A graph candidate is blamed if (a) a correct process is not
  in the graph or (b) a process receives an out of date message
  through one of its incoming links.  In both cases the candidate is
  definitively removed from the $Candidates$ sets of all processes.
  To achieve this goal the process sends an accusation message ($ACC$)
  using a reliable broadcast and uses an array $Heard$ that ensures that an
  identical candidate (that is, the same graph with the same
  accusation and proposition values) can never be added again.
  Moreover, upon delivery of an accusation message for graph $x$,
  $root[x]$ increments $Acc[x]$.
\end{itemize}

We now present different mechanisms used to obtain the efficiency.


For all graphs $x \in \X$, only the process $root(x)$ is allowed to
propose $x$ as a candidate to the rest.
 Each process $p$ stores its
better candidate in its variable $me$, that is, the least blamed graph
$x$ such that $root(x) = p$.
\begin{itemize}
\item If a process finds in $Candidates$ a better candidate than $me$,
  it removes $me$ from $Candidates$. 
\item If a process finds that $me$ is better, it adds $me$ to
  $Candidates$ and sends a $new$ message containing $me$ (1) to all
  processes that are not in $Node(me)$, and (2) to immediate
  successors of $p$ in $me$.  The immediate successors in $me$ add
  $me$ to their $Candidates$ set and relay the $new$ message, and so
  on.  By the reliability of the links, every correct process that is
  not in $me$ eventually receives this message and blames $me$.
  
\end{itemize}
These mechanisms are achieved by the procedure
$updateExtractedGraph()$. This procedure is called each time a graph
candidate is blamed or a new candidate is proposed.  Note that the
$Candidates$ set is maintained with the set $OtherCand$ (the
candidates of other processes), a boolean $Local$ that is true when
the process has a candidate, and $me$, the graph candidate.


A process $p$ may give up a candidate without this candidate being
blamed: in this case, $p$ is the root of the candidate, it finds a
better candidate in $OtherCand$, and removes $me$ from
$Candidates$. Then, $p$ must not increment $Acc[me]$ when it receives
accusations caused by this removing, indeed these accusations are not
due to delayed messages. That is the goal of the proposition counter
($Prop$): in $Prop[x]$, $root(x)$ counts the number of times it
proposes $x$ as candidate and includes this value in each of its $new$
messages (to inform other process of the current value of the
counter). Hence, when $q$ wants to blame $x$, it now includes its own
view of $Prop[x]$ in the accusation message. This accusation will be
considered as legitimate by $root[x]$ (that is, will cause an
increment of $Acc[x]$) only when the proposition counter inside the
message matches $Prop[x]$. Also, whenever $root[x]$ removes $x$ from
$Candidates$, $root[x]$ increments $Prop[x]$ and does not send the new
value to the other processes. In this way accusations due to this
removing will be ignored.

For any timely candidate, the accusation counter will be bounded and
its proposition counter increased each time it is proposed.  In this
way the graph with the smallest accusation and proposition values
eventually remains forever in the $Candidates$ set of all correct
processes and it is chosen as extracted graph.  (This is done in the
procedure $updateExtracted Graph()$.) Moreover, eventually all other
candidates are given up and it remains only this graph in $Candidates$.
In this way, only  $alive$ messages are sent and they are
sent along the directed edges of the extracted graph ensuring the
efficiency.

\begin{figure}[htpb]
\scriptsize

{\tt Code for each process $p$}
\begin{algorithmic}[1]

\State {\bf Procedure} $updateExtractedGraph()$
\State \qquad Let $(a_{min},min) = \min_{\prec_{lex}} \{(acc,c)$ such that $(c,acc,-,-) \in OtherCand\} \cup \{(\infty,\infty)\}$
\State \qquad {\bf if} {$(a_{min},min)<( Acc[me],me) \wedge Local$} {\bf then} \Rem{Give up $me$}
	\State \qquad \qquad \brd{$ACC$,$me$,$Acc[me]$,$Prop[me]$,$\Delta[me]$}
	\State \qquad \qquad $Prop[me] \gets Prop[me] + 1$ 	
	\State \qquad \qquad $Local \gets false$
\State \qquad {$Candidates \gets OtherCand$}

\State \qquad $me \gets x$ such that $(a,x) = \min_{\prec_{lex}} \{(acc,c)$ such that $c \in \X \wedge root(c)=p\}$ 
\State \qquad {\bf if} {$(Acc[me],me)<(a_{min},min) \wedge Local = false$} {\bf then}         \Rem{Propose $me$}
	\State \qquad \qquad $Local \gets true$
	
	\State \qquad \qquad {$Candidates \gets Candidates \cup \{(me,Acc[me],Prop[me],\Delta[me])\}$}
         \State \qquad \qquad \send{$new$,$me$,$Acc[me]$,$Prop[me]$,$\Delta[me]$}{every process not in $Node(me)$}  

	\State \qquad \qquad {\bf for all} {$h \in \Pi \setminus \{p\}$} {\bf do}
        \State \qquad \qquad \qquad {\bf if} {($h$,$p$)$\in Edge(me)$} {\bf then}
        \State \qquad \qquad \qquad \qquad $\Delta[h] $$\gets$ $\max(\Delta[h],\Delta[me])$
        \State \qquad \qquad \qquad \qquad \settimer{$h$} $\gets$ $\Delta[h]$
          \State   \qquad \qquad  \qquad {\bf if} {($p$,$h$)$\in Edge(me)$ and $h \neq root(me)$} {\bf then}
             \State \qquad \qquad \qquad \qquad{\send{$new$,$me,Acc[me],Prop[me],\Delta[me]$}{$h$}}
  
\State \qquad $G \gets x$ such that $(a,x) \min_{\prec_{lex}}\{(a',x')$ such that $(x',a',p',d') \in Candidates\}$


\algstore{store}
\end{algorithmic}
\caption{Procedure  updateExtractedGraph of Algorithm ${\cal AF}({\X})$}\label{algo:eff2}
\end{figure}

\begin{figure}[t!]
\scriptsize
{\tt Code for each process $p$}
\begin{algorithmic}[1]
\algrestore{store}
\State On initialization:

\State {\bf for all} {$x \in \X$ such that $root(x) = p$} {\bf do}
\State \qquad $Acc[x] \gets 0$;  $Prop[x] \gets 0$; $\Delta[x] \gets n $ 
\State {\bf for all} $x \in \X $ such that $root(x)\neq p$ {\bf do} $Heard[x] \gets (-1,-1)$

\State {\bf for all} {$q \in \Pi \setminus \{p\}$} {\bf do} $\Delta[q] \gets 1$
\State $OtherCand \gets \emptyset$
\State $Local \gets false$
\State $me \gets  min\{x$ such that $x \in \X \wedge root(x)=p\}$
\State  $updateExtractedGraph()$
\State  {\bf start tasks} 1 and  2

\vspace{0.5em}

\State {task 1:}
\State  \qquad {\bf loop forever}
\State \qquad \qquad \  \send{$alive$}{every process $q$ such that $\exists (x,$-$,$-$,$-$)$$\in Candidates$ and $(p,q)\in Edge(x)$ {\bf every $K$ time}}  

\vspace{0.5em}

\State {task 2:}

\State \qquad  {\bf upon} {\receive{$alive$}{$q$}} {\bf do} 
\State \qquad \qquad  $\settimer{q} \gets$ $\Delta[q]$

\vspace{0.5em}

\State \qquad  {\bf upon} {\expire{$q$}}  {\bf do}   \Rem{Link $(q,p)$ is not timely, blame all candidates that contains $(q,p)$}
\State \qquad \qquad {\bf for all} {$(x,a,pr,d) \in OtherCand$ such that $(q,p) \in Edge(x)$} {\bf do}

\State \qquad  \qquad \qquad \brd{$ACC$,$x$,$a$,$pr$,$d$}

\State \qquad  \qquad {\bf if} {$(q,p) \in Edge(me)$} {\bf then}
\State \qquad  \qquad \qquad \brd{$ACC$,$me$,$Acc[me]$,$Prop[me]$,$\Delta[me]$}
\vspace{0.5em}

\State \qquad {\bf upon} {\receive{$new, x ,a, pr, d$}{$q$}} {\bf do}\Rem{Proposition of a new  candidate}
	\State \qquad  \qquad  {\bf if} {$p \notin Node(x)$} {\bf then} \Rem{Blame $x$ that does not contain $p$}
	\State \qquad  \qquad \qquad  \brd{$ACC$,$x$,$a$,$pr$}
	\State \qquad  \qquad  {\bf else} 
		\State   \qquad \qquad \qquad $newCand  \gets false$ 
		\State   \qquad \qquad \qquad {\bf if}  {$ (x,-,-,-) \notin OtherCand$ and $Heard(x)<(a,pr)$} {\bf then} \Rem{New  candidate}
  
                \State   \qquad \qquad \qquad \qquad $newCand  \gets true$ 

		\State  \qquad \qquad \qquad {\bf if} {$\exists (x,a_c,pr_c,d_c) \in OtherCand$ with $(a_c,pr_c)<(a,pr)$} {\bf then} \Rem{New  candidate}
		\State   \qquad  \qquad \qquad  \qquad $OtherCand \gets OtherCand \setminus (c,a_c,pr_c,d_c)$
                \State   \qquad \qquad   \qquad \qquad $newCand  \gets true$ 

		\State   \qquad  \qquad  \qquad {\bf if} {$ newCand$ } {\bf then} 
		\State  \qquad \qquad \qquad \qquad $OtherCand \gets OtherCand\cup (x,a,pr,d)$
		\State  \qquad \qquad \qquad \qquad $updateExtractedGraph()$
		\State   \qquad \qquad   \qquad \qquad $Heard[x] \gets (a,pr)$
		\State  \qquad \qquad  \qquad  \qquad {\bf for all} {$h \in \Pi \setminus \{p\}$} {\bf do}
			\State  \qquad \qquad  \qquad  \qquad \qquad {\bf if} {($h$,$p$)$\in Edge(x)$} {\bf then} 
			\State   \qquad \qquad \qquad \qquad  \qquad \qquad $\Delta[h] $$\gets$ $\max(\Delta[h],d)$
			\State   \qquad \qquad \qquad \qquad \qquad  \qquad \settimer{$h$}$ \gets \Delta[h]$	
			\State   \qquad \qquad \qquad    \qquad \qquad {\bf if} {($p$,$h$)$\in Edge(x)$ and $h \neq root(x)$} {\bf then} {\send{$new, x ,a, pr, d$}{$h$}}
		\vspace{0.5em}

\State \qquad {\bf upon} {\deliv{$ACC$,$x$,$a$,$pr$,$d$}} {\bf do}
	\State \qquad \qquad {\bf if} {$root(x)=p$} {\bf then}
		\State \qquad \qquad \qquad {\bf if} {$x=me \wedge a = Acc[me] \wedge pr = Prop[me]$} {\bf then} \Rem{Check if the accusation is up to date}
		\State \qquad \qquad \qquad \qquad $Acc[me] \gets Acc[me] + 1$; $\Delta[me] \gets \Delta[me] + 1$
		\State \qquad \qquad \qquad \qquad $Local  \gets false$
	\State \qquad \qquad {\bf else} 
	\State \qquad \qquad \qquad $OtherCand \gets OtherCand \setminus (x,a,pr,d)$
	\State \qquad \qquad \qquad {\bf if} {$ Heard[x] <(a,pr)$} {\bf then} $Heard[x]\gets (a,pr)$
	\State \qquad \qquad  $updateExtractedGraph()$

\end{algorithmic}
\caption{Algorithm ${\cal AF}({\X})$ that efficiently extracts a graph in  $\X$ }\label{algo:eff}
\end{figure}

A sketch of the correctness proof of ${\cal AF}({\X})$ is given in the appendix.
Then, we can conclude with the following theorem:

\begin{theorem}\label{th:effCN}
  Let $\X$ be a dicut-closed system with a root.  Algorithm ${\cal
    A}({\X})$ efficiently extracts a graph
  in $\X$.  Moreover if all graphs of $\X$ are strongly connected,
  Algorithm ${\cal A}({\X})$ efficiently and
  exactly extracts a graph in $\X$.
\end{theorem}

\section{Conclusion}\label{sect:ccl}
Failure detector implementations in partially synchronous models
generally use the timeliness properties of the system to approximate
the set of correct (or faulty) processes. In some way, the extraction problem
is a kind of generalization: instead of only searching the set of
correct processes, here we try to extract also information about the timeliness
of links. 
Besides, our solutions
are based on already existing mechanisms used in failure detectors
implementations as in~\cite{ADFT03,ADFT04}.



Information about the timeliness of links is useful for efficienecy of
fault-tolerant algorithms. 
In particular,
in any extracted graph,  any path between a pair
of correct processes is only constituted of timely links. This
property is particulary interesting to get efficient routing algorithms.

We gave an extraction algorithm for  dicut-closed set of 
timeliness graphs.  Moreover, we proved that the extraction is exact
when all the  timeliness graphs are also strongly connected.

Given  dicut-closed timeliness graphs that contains a root, we shown
how to efficiently extract a graph from it. By efficiency we mean
giving a solution where eventually messages are only sent over the
links of the extracted graph.

It is important to note that the main purpose of the algorithms we
proposed is to show the feasability of the extraction under some
conditions. So, the complexity of our algorithms was not the main
focus of this paper.

As a consequence, our algorithms are somehow unrealistic because of
their high complexity. Giving more practical solutions will be the
purpose of our future works.

\subsection*{Acknowledgments} We are grateful to members of the {\em GRAPH} team of the {\em LIAFA} Lab for the helpful discussions and their interesting suggestions.




{\footnotesize
\bibliographystyle{plain}
\bibliography{biblio.bib}
}


 \pagebreak \small

 \begin{appendix}

\section{Appendix}
\subsection{Proof of Proposition~\ref{prop:noalgo}}

\noindent{\bf Proposition~\ref{prop:noalgo}}
  {\em 
  There exists some systems $\cal X$ for which there
  is no extraction algorithm.}

\bigskip

\begin{sketch}


  Assume there is an extraction algorithm $\cal A$ for  $\cal
  PAIR$ with 5 processes.

  Consider a run $r$ of $\cal A$ in system $\cal PAIR$ with
  $T(r)=\langle \{p_1,p_2,p_3,p_4,p_5\},$ $ \{(p_1,p_2) ,(p_2,p_1),
  (p_3,p_4), $ $(p_4,p_3) \}\rangle$.  To satisfy the properties of
  the extraction, $\langle \{p_1,p_2,p_3,p_4,p_5\},$ $ \{(p_1,p_2),$
  $(p_2,p_1) \}\rangle$ or $\langle \{p_1,p_2,p_3,p_4,p_5\},$ $
  \{(p_3,p_4),$ $(p_4,p_3) \}\rangle$ must be extracted from the run
  $r$. There is a time $t_1$ after which $r$ converges for example to
  $\langle\{p_1,p_2,p_3,p_4,p_5\},$ $ \{(p_1,p_2) ,(p_2,p_1) \}\rangle$.

  Consider now run $r'$ of $\cal A$ in system $\cal PAIR$ with
  $T(r')=\langle\{p_3,p_4,p_5\},$ $ \{ (p_3,p_4), (p_4,p_3) \}\rangle$
  such that $r$ and $r'$ are indistinguishable until time $t_1$ and
  $p_1$ and $p_2$ crash in $r'$ at time $t_1+1$.  There is a time
  $t_2$ after which $r'$ converges to a graph with the directed edges
  $\{ (p_3,p_4) ,(p_4,p_3) \}$.
	
  Consider now that in $r$ all messages from $p_1$ and $p_2$ to
  $\{p_3,p_4,p_5\}$ sent after time $t_1$ are delayed after time $t_2$. For
  $p_5$, the runs $r$ and $r'$ are indistinguishable until $t_2$.  So,
  at time $t_2$, $p_5$ outputs a graph with directed edges $\{
  (p_3,p_4), (p_4,p_3) \}$.
	  
  Now consider run $r''$ of $\cal A$ in system $\cal PAIR$ with
  $T(r'')=\langle\{p_1,p_2,p_5\},$ $ \{(p_1,p_2) ,(p_2,p_1) \}\rangle$
  such that $r$ and $r''$ are indistinguishable until time $t_2$ and
  $p_3$ and $p_4$ crash in $r''$ at time $t_2+1$.  There is a time
  $t_3$ after which $r''$ converges to a graph with the directed edges
  $\{(p_1,p_2) ,(p_2,p_1) \}$.

  Consider again that in the run $r$ all messages from $p_3$ and $p_4$
  to $\{p_1,p_2,p_5\}$ sent after time $t_2$ are delayed after
  $t_3$. For $p_5$ the runs $r$ and $r''$ are indistinguishable.  So,
  at time $t_3$, $p_5$ outputs a graph with directed edges
  $\{(p_1,p_2) ,(p_2,p_1) \}$.

  Inductively, we can construct the run $r$ in such a way that $p_5$
  alternates forever between a graph with directed edges $\{(p_1,p_2)
  ,(p_2,p_1) \}$ and a graph with directed edges $\{(p_3,p_4)
  ,(p_4,p_3) \}$ and never converges definitively.  This contradicts
  the existence of an algorithm that extracts a graph in $\cal PAIR$.
\end{sketch}

\subsection{Proof of Proposition~\ref{prop:noexact}}

\noindent {\bf Proposition~\ref{prop:noexact}}
{\em There exists some systems $\cal X$ for which there
  is an  extraction algorithm and there is no exact extraction algorithm.}

\bigskip

\begin{sketch}
Consider the system  $\cal
  TREE$ with 3 processes.
We prove in the next section that there is an extraction algorithm for  this system.
  Assume there is an {\em exact} extraction algorithm $\cal A$ for this system.
  
  Consider a run $r$ of $\cal A$ in this system with
  $T(r)=\langle \{p_1,p_2,p_3\},$ $ \{(p_1,p_2) ,(p_1,p_3)\}\rangle$.  To satisfy the properties of the
 exact extraction, there is a time $t_1$ after which the graph $\langle
  \{p_1,p_2,p_3\},$ $ \{(p_1,p_2),$ $(p_1,p_3) \}\rangle$ is extracted.

  Consider now run $r'$ of $\cal A$ in system $\cal TREE$ with
  $T(r')=\langle\{p_1,p_2\},$ $ \{ (p_1,p_2)\}\rangle$ such that $r$
  and $r'$ are indistinguishable until time $t_1$ and $p_3$ crashes in
  $r'$ at time $t_1+1$.  There is a time $t_2$ after which $r'$
  converges to $\langle \{p_1,p_2\}, \{(p_1,p_2) \}\rangle$ .
  	
  Consider now that in $r$ all messages from $p_3$ to $\{p_1, p_2\}$ 
sent after time $t_1$ are delayed after time $t_2$. For
  $p_1$, the run $r$ and $r'$ are indistinguishable until $t_2$.  So,
  at time $t_2$, $p_1$ outputs  $\langle \{p_1,p_2\}, \{ (p_1,p_2)  \}\rangle$.

  Inductively, we can construct the run $r$ in such a way that $p_1$
  alternates forever between a graph $\langle \{p_1,p_2,p_3\},$ $
  \{(p_1,p_2),$ $(p_1,p_3) \}\rangle$ and a graph $\langle
  \{p_1,p_2\},$ $ \{(p_1,p_2) \}\rangle$ and never converges
  definitively.  This contradicts the existence of an algorithm that
  exactly extracts a graph in $\cal TREE$.
\end{sketch}

\subsection{Proof of Theorem ~\ref{th:effCN}}
In this section, we propose a sketch of the correctness proof of the efficient extraction algorithm ${\cal
  AF}({\X})$ (Figures \ref{algo:eff2} and \ref{algo:eff}). 
  In this sketch, we consider a run $r$ of ${\cal AF}({\X})$ in
dicut-closed system with a root, $\cal X$. We will denote by
$var_p^t$ the value of $var_p$ at time $t$.

We first notice that all variables $Acc[x]$ and $Prop[x]$ can only be
modified by the process $root(x)$ and are increasing:

\begin{lemma}
  For all time $t$ and $t'$, $t \ge t'$, for all processes $p$, for
  all graphs $x$ in $\X$ such that $p=root(x)$, $Acc^t_p[x] \ge
  Acc^{t'}_p[x] $ and $Prop^t_p[x] \ge Prop^{t'}_p[x]$.
\end{lemma}

Consider a graph $x$ such that its root $p$ crashes. Eventually,
every process $q$ such that $x \in OtherCand$ and $(p,q) \in Edge(x)$
reliably broadcasts an accusation for $x$. This way, $x$ is removed
from the $OtherCand$ set of any correct process and never more added
(because $p$ is crashed), hence:

\begin{lemma}
  If $p$ is faulty, there exists a time $t$ such that for all graphs
  $x$ of $\X$ with $root(x)=p$, for all correct processes $q$ in
  $r$, for all $ t'\ge t$: $x \notin OtherCand^{t'}_q$.
\end{lemma}

As $r$ is a run of $\X$, there exists some timeliness graph $o$ in
\cal $\X$ such that $o$ is compatible with $T[r]$. In this case,
$Nodes(o)=Correct(r)$ and the process $root(o)$ is a correct
process:

\begin{lemma} \label{eff:exist}
  There exists a timeliness graph $o$ of $\X$ such that $o$ is
  compatible with $T(r)$ and $root(o)$ is a correct process.
\end{lemma}

Moreover:

\begin{lemma}\label{eff:bound}
  Let $o$ be a timeliness graph of $\X$ such that $o[Correct(r)]$ is a
  compatible with $T(r)$ and $root(o)$ is a correct process:
  $Acc_{root(o)}[o]$ is bounded.
\end{lemma}

For all correct processes $p$, for all graphs $x$ in $\X$ with
$root(x)=p$, let $A[x]_p$ be the largest value of $Acc[x]_p$ in $r$
($\infty$ if $Acc[x]_p$ is unbounded).  Let $g$ to be the graph with
the smallest $A[g]_p$
(break ties by the total order on
graphs). Let $C$ be the value of $A[g]_p$.

Note that from Lemma~\ref{eff:exist} and  Lemma~\ref{eff:bound}, $C <
\infty$. Moreover, by construction of $g$, $root(g)$ is a correct
process, $root(g)$ eventually elects $g$ forever
($me_{root(g)}=g$), and as a consequence $Prop[g]_{root(g)}$
becomes constant:


\begin{lemma}\label{effp1}
There exists a time after which   $me_{root(g)}=g$.
\end{lemma}
\begin{lemma}\label{effp2}
There exists a time after which   $Prop[g]_{root(g)}$ stops changing.
\end{lemma}

 Let $P$ be the largest  value of the proposition counter of
$g$ ($Prop[g]$).
The following three lemmas are immediate consequences of Lemma
\ref{effp1}:

\begin{lemma}\label{effp4}
  For every correct process $p\neq root(g)$, there exists a time
  after which $g \in OtherCand_p$.
\end{lemma}

\begin{lemma}\label{effp3}
  There exists a time after which $me_{root(g)}=g$ and
  $Local_{root(g)}= true$ and $OtherCand_{root(g)}=\emptyset$.
\end{lemma}

\begin{lemma}\label{effp5}
  For every correct process $p\neq root(g)$, there exists a time
  after which $OtherCand_p=\{g\}$ and $Local_p=false$.
\end{lemma}

From Lemmas~\ref{effp3} and~\ref{effp5}, the algorithm converges to a
graph of $\X$:

\begin{lemma}\label{effterm}
  There exists a timeliness graph $x \in \X$ (actually $g$) such that
  every correct process $q$ outputs $x$ forever.
\end{lemma}

From Lemma~\ref{effp3} and Lemma~\ref{effp5}, we can deduce that the
algorithm is efficient:

\begin{lemma}\label{effeff}
  There is a time after which every correct process $p$ sends messages
  only to the process $q$ such that there is a directed edge $(p,q)$
  in $Edge(g)$.
\end{lemma}

From the Lemma~\ref{effterm}, we deduce  the Convergence and the
Validity properties.

It remains to prove that $g$ satisfies the properties of the
approximation: (1) $g[Correct(r)]$ is compatible with $T[r]$, and (2)
$g[Correct(r)]$ is a dicut reduction of $g$ or is equal to $g$.

When $root(g)$ sets $Local$ to true and $me$ to $(g,C,P,-) $, it
sends a message $new$ to all processes (recall that  $C$ the final value of the
accusation counter of $g$  and $P$   the final value of its the proposition counter.).  
As the links are reliable, all
correct processes eventually receives this message. If a correct
process $q$ is not in $Node(g)$, it reliably broadcasts an accusation
message $ACC$.  When process $root(g)$ delivers such a broadcast, it
increments the accusation counter of $g$ contradicting the fact that
$Acc[g]$ is bounded by $C$, hence:

\begin{lemma}\label{effl1}
$Correct(r) \subseteq Node(g)$.
\end{lemma}

When a correct process receives this $new$ message, it sends $\langle
alive\rangle$ to every process $q$ such that $(p,q)$ in $Edge(g)$.
And it monitors all incoming links $(q,p)$ such that $(q,p)$ in
$Edge(g)$.  If there is a link $(a,b)$ of $ Edge(g)$ between two
correct processes $a$ and $b$, then $a$ sends regularly $alive$
message to $b$.  By construction of $g$, $b$ never blames $g$, then
$b$ receives no out of date message. By the FIFO property of the link,
the link is timely:

\begin{lemma}\label{eff:compat}
$g[Correct(r)] $ is compatible with  $T[r]$.
\end{lemma}

By Lemma~\ref{effl1}, $Node(g) =Correct(r)\cup F$.

If $F$ is empty the Closure property is trivially ensured.  We now
consider the case where $F$ is not empty. $F$ contains only faulty
processes.  If there is an edge in $Edge(g)$ from a faulty process $q$
to a correct process $p$, eventually the process $p$ stops receiving
messages from $q$ and the accusation counter of $g$ will be
incremented, which contradicts the fact that the accusation counter of
$g$ remains equal to $C$ forever. So we have:

\begin{lemma}\label{eff:dicutclose}
If  $F \neq \emptyset$ then  $Edge(g) \cap( F\times Correct(r))=\emptyset$.
\end{lemma}

We showed the Convergence (Lemma~\ref{effterm}), the Validity
(Lemma~\ref{effterm}), the Compatibility (Lemma~\ref{eff:compat}),
the closure (Lemma~\ref{eff:dicutclose}), and the Efficiency (Lemma~\ref{effeff}). Moreover,
Proposition~\ref{prop:1} shows the exact extraction when all graphs of
$\X$ are strongly connected. Hence, we can conclude with the following
theorem:

\vspace{0.5em}

\noindent{\bf Theorem~\ref{th:effCN}}
  {\em Let $\X$ be a dicut-closed system with a root.  Algorithm ${\cal
    A}({\X})$ efficiently extracts a graph
  in $\X$.  Moreover if all graphs of $\X$ are strongly connected,
  Algorithm ${\cal A}({\X})$ efficiently and
  exactly extracts a graph in $\X$.}

 \end{appendix}

\end{document}